\documentclass{aa}  
\usepackage{graphicx}
\usepackage{txfonts}

\usepackage{epsfig,psfig}
\usepackage{array,longtable}

\newcommand{\E}[1]{\times 10^{#1}}
\newcommand{\s}{\,{\rm s}}      \newcommand{\ps}{\,{\rm s}$^{-1}$}
    \newcommand{\Msun}{{\rm M}_{\rm \odot}}
\newcommand{\cm}{\,{\rm cm}}    \newcommand{\km}{\,{\rm km}}

\newcommand{\ncol}{$N({\rm H_2})$} \newcommand{\tex}{$T_{\rm ex}$}
\newcommand{\brightnessunit}{W~m$^{-2}$~Hz$^{-1}$~sr$^{-1}$}

\newcommand{\du}{$d_{3.2}$}

\newcommand{\HI}{H~{\sc i}}
\newcommand{\HII}{H~{\sc ii}}
\newcommand{\vlsr}{$V_{\rm LSR}$}       \newcommand{\tmb}{$T_{\rm mb}$}
\newcommand{\twCO}{$^{12}$CO}  \newcommand{\thCO}{$^{13}$CO}
\newcommand{\CeiO}{C$^{18}$O}
\newcommand{\snr}{{G16.11$-$0.51}}

\titlerunning{SNR G16.11$-$0.51}
\authorrunning{Zhou et al.}

\begin{document}

\title{Discovery of an old supernova remnant candidate through carbon monoxide line emission}

\author{Xin Zhou\inst{1}, Yang Su\inst{1}, Ji Yang\inst{1}, Yang Chen\inst{2,3}, \and Zhibo Jiang\inst{1}
}

\institute{Purple Mountain Observatory and Key Laboratory of Radio Astronomy, Chinese Academy of Sciences, 10 Yuanhua Road, Nanjing 210023, China\\ 
\email{xinzhou@pmo.ac.cn}
\and
School of Astronomy \& Space Science, Nanjing University, 163 Xianlin Avenue, Nanjing 210023, China
\and
Key Laboratory of Modern Astronomy and Astrophysics, Nanjing University, Ministry of Education, Nanjing 210023, China}


 
  \abstract
{Most old supernova remnants (SNRs) in the Milky Way have not yet been identified. Considering their large potential number and the sufficient momentum-energy transfer to the interstellar medium (ISM), they are a key part of our understanding of the overall role of SNRs in the ISM.
Here we report our discovery of an expanding molecular shell identified by CO line observations, namely \snr. It covers a known SNR, specifically G16.0$-$0.5, and is larger in size, i.e.\ 0.56$^\circ$ over 0.20$^\circ$.
Based on its spatial and kinematic structures, weak nonthermal radio-continuum emission, and derived physical properties, we suggest that it is an old SNR.
At a systemic velocity of +41.3 \km\ps, the best estimated kinematic distance of \snr\ is $\sim$3.2 kpc, implying its radius of about 15.6~pc.
The age of \snr\ is estimated to be greater than $\sim$$10^{5}$~yr, and, in a dense molecular environment, it has formed dense and thin shell layers. The kinetic energy of the expanding molecular gas of \snr\ is about $6.4\E{49}$~erg, accounting for approximately six percent of the initial SN explosion energy.
Although old SNRs have essentially become cold and hard to detect, our discovery suggests that they can be found by searching for CO line emissions.}



   \maketitle
%

\section{Introduction}
Supernova remnants (SNRs) contain a large amount of momentum, energy, and heavy elements. They sufficiently transfer them to the surrounding interstellar medium (ISM) and cool down in their late stages \citep[e.g., $>\sim10^5$;][]{Koo+2020}.
One supernova event occurs on average every $40\pm10$ yr in the Milky Way \citep{Tammann+1994}, resulting in over 2000 SNRs younger than $10^5$ yr.
If we take $10^6$ yr to be the lifetime of an SNR, there would be over 20000 SNRs in total.
The number of old SNRs is much larger than that of young bright ones.
Old SNRs are a key part of understanding the overall role of SNRs in the ISM, especially their impact on molecular clouds (MCs; e.g., how they regulate the formation of next generation stars).
Old SNRs that have a large potential number and large size are also likely to overlap with other sources, and they act as backgrounds to affect our understanding of other sources.

Only a limited number of SNRs have been identified in our Galaxy, i.e.\ fewer than 400, mostly based on their radio-continuum emission \citep[cf.][]{Green2019, FerrandSafi-Harb2012}.
Most of these known SNRs are bright and young. 
It is difficult to detect old SNRs because they usually have very weak radiation; for example, they are radio faint and begin to dissolve into the ISM.
Benefiting from enhanced observations in different wavelengths, new SNR candidates have been continuously discovered, e.g., by optical emission \citep{Fesen+2020}, by X-ray emission \citep{Becker+2021, Churazov+2021, Khabibullin+2023}, and mostly by radio-continuum emission \citep[][etc.]{Anderson+2017, Hurley-Walker+2019c, Gao+2020, Dokara+2021, Ball+2023}.
It appears that there are still a large number of potential SNRs waiting to be discovered; nevertheless, background emission would limit such detections \citep[see, for example, radio-continuum emission from HESS J1912+101 confused by background diffuse emission][]{ReichSun2019}, especially for weak SNRs or toward the inner Galactic region.
Additionally, the large size and quantity of old SNRs can lead to a severe overlapping effect, further complicating the situation.
Considering that old SNRs would form a cold, dense, and gradually momentum-conserving shell, \HI\ 21~cm line emission has been used as a good tracer to detect such old SNRs, especially large ones at high Galactic latitude or with localized high-velocity \HI\ features \citep[e.g.,][etc.]{KooKang2004, Koo+2006, KangKoo2007, Kang+2012, XiaoZhu2014}.
Although CO line emission can also trace dense shells associated with SNRs \citep[e.g., for known SNRs;][]{Chen+2014, Sofue+2021}, no known SNR has been discovered through CO lines to date \citep{Green2019, FerrandSafi-Harb2012}.
Note that, based on the analysis of CO and \HI\ data, \cite{Su+2017b} suggested the SNR origin of the TeV source HESS J1912+101.
Compared to \HI\ gas, molecular gases are more dense and clumpy, hence, it is difficult to detect a large complete shell-like structure associated with an SNR.
In addition, molecular lines are very likely to have serious overlapping effects too, especially toward the inner Galactic region, which can complicate their velocity features.
Nevertheless, since remnants of core-collapse supernovae are thought to be located close to their parent MCs, numerous molecular shells are expected to have originated from old SNRs, e.g., some CO bubbles in the W43 molecular complex were interpreted as fully evolved SNRs by \cite{Sofue2021}.

Here we report the discovery of a new expanding CO shell, namely \snr, which covers a known SNR, i.e.\ G16.0$-$0.5. 
Its spatial and kinematic structures, weak nonthermal radio-continuum emission, and derived physical properties indicate its origin from an SN explosion.

\section{Observations}\label{sec:obs}
The CO line emission was observed during November 2011 to November 2013 and from February to April 2019 using the Purple Mountain Observatory (PMO) 13.7~m millimeter-wavelength telescope located in Delingha, China, as part of the Milky Way Imaging Scroll Painting (MWISP) project\footnote{http://english.dlh.pmo.cas.cn/ic/}. 
The $3\times3$ multibeam sideband separation receiver, i.e.\ Superconducting Spectroscopic Array Receiver \citep[SSAR,][]{Shan+2012}, was used to simultaneously observe the \twCO~(J=1--0), \thCO~(J=1--0), and \CeiO~(J=1--0)  lines.
The fast Fourier transform (FFT) spectrometer with 1 GHz bandwidth and 16384 channels was used as the backend for each sideband.
Correspondingly, the spectral resolutions of the three CO lines were 0.17~\km\ps\ for \twCO~(J=1--0) and 0.16~\km\ps\ for both \thCO~(J=1--0) and \CeiO~(J=1--0).
We mapped a $1.2^\circ\times 1.2^\circ$ area via on-the-fly (OTF) observing mode.
Following the standard procedure of the MWISP CO line survey, observations were performed at least twice along the Galactic longitude and latitude directions, respectively, and specialized data checks were carried out. Data with bad baselines were excluded, and thereby baseline problems were largely suppressed. However, some minor rms noise inhomogeneities were introduced, for example, along the Galactic longitude or latitude direction.
The total error in pointing and tracking was within $5''$.
The data were meshed with a grid spacing of $30''$., and the half-power beam width (HPBW) was $\sim51"$.
The typical main-beam efficiencies $\eta_{\rm mb}$ were $\sim$48\% for USB and $\sim$52\% for LSB, which was used to convert the antenna temperature to the main beam temperature by $T_{\rm mb}=T_A^*/\eta_{\rm mb}$ (see details in the status report of the telescope\footnote{http://www.radioast.nsdc.cn/mwisp.php}).
After performing the linear baseline subtraction process, the typical rms noises obtained were $\sim0.43$~K per channel for \twCO~(J=1--0) and $\sim0.21$~K per channel for \thCO~(J=1--0) and \CeiO~(J=1--0).
All data were reduced using the GILDAS/CLASS package\footnote{http://www.iram.fr/IRAMFR/GILDAS}.

Radio continuum emission data in four bands with frequencies ranging from 72 to 103 MHz, 103 to 134 MHz, 138 to 170 MHz, and 170 to 231 MHz were obtained from the Galactic and Extragalactic All-sky Murchison Widefield Array (GLEAM) survey\footnote{https://www.mwatelescope.org/gleam} \citep{Wayth+2015, Hurley-Walker+2017, Hurley-Walker+2019a}.
Other radio-continuum emission data in six bands were obtained from The HI/OH/Recombination line survey \citep[THOR;][]{Beuther+2016, Wang+2020}, which are centered at 1.06, 1.31, 1.44, 1.69, 1.82, and 1.95 GHz and have a bandwidth of 128 MHz.
Combined 1.4 GHz continuum data from the THOR and the VLA Galactic Plane Survey \citep[VGPS,][]{Stil+2006} was also applied, from which the flux retrieved is consistent with the literature \citep{Anderson+2017, Wang+2020}.

\section{Results}\label{sec:result}

\begin{figure*}[ptbh!]
\centerline{{\hfil\hfil
\psfig{figure=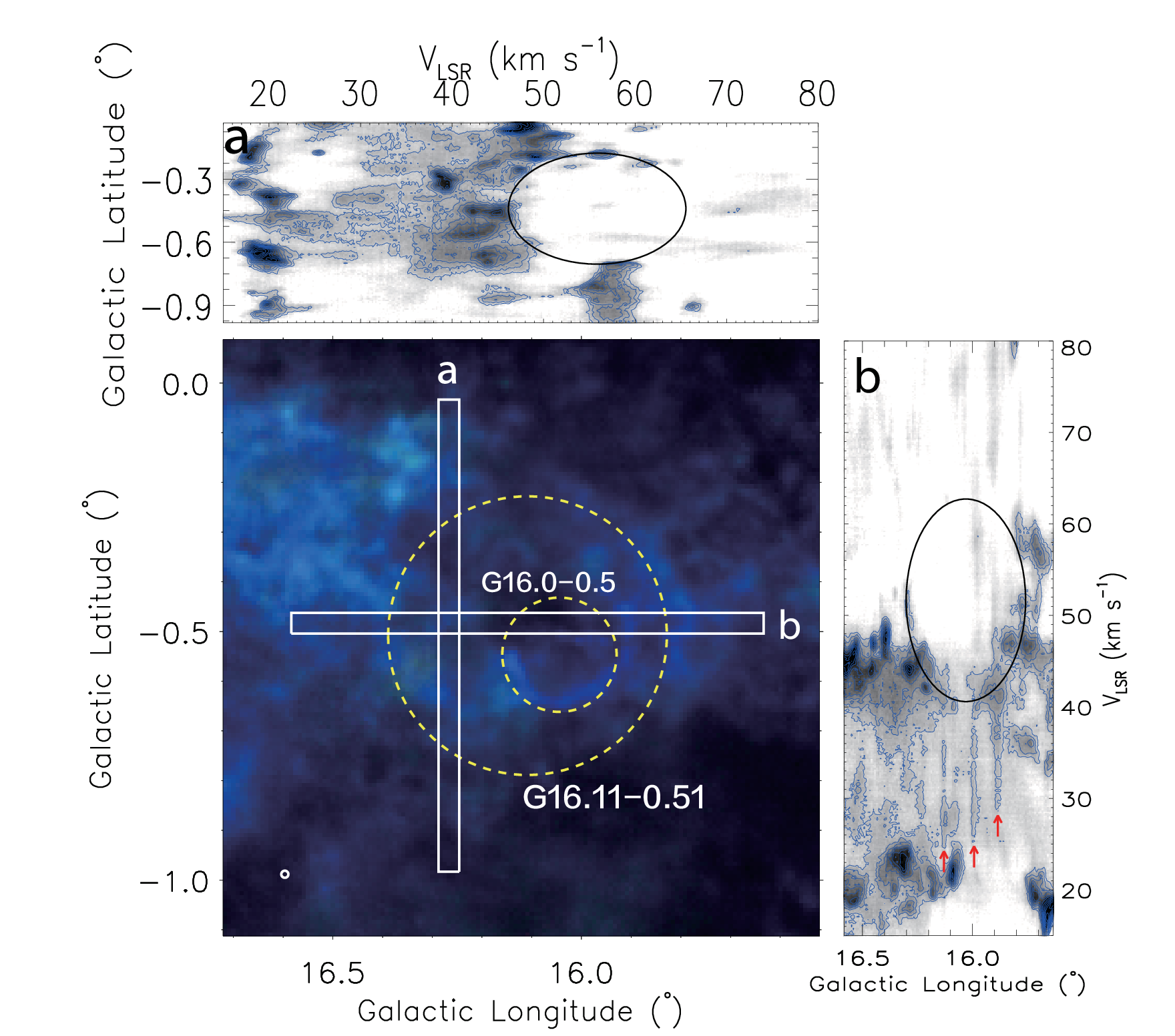,height=5.5in,angle=0, clip=}
\hfil\hfil}}
\caption{Integrated intensity and position-velocity maps of CO emission toward \snr. {\it Bottom left panel:} pseudo tricolor image of the integrated intensity of \twCO\ (J=1--0) (blue), \thCO\ (J=1--0) (green), and \CeiO (J=1--0) (red) line emission in the velocity range from +35 to +85 \km\ps. The intensities of all three CO lines have confidence levels over $5\sigma$. The large and small yellow dashed circles indicate the extent of \snr\ and SNR G16.0$-$0.5, respectively.
The beam is represented by a white circle in the lower left corner. 
The top and right panels show the \twCO\ (J=1--0) position-velocity maps along the strips indicated in the intensity map. The minimum value of the background grayscale map is $1\sigma$, and the contour levels are from $3\sigma$ and in a step of $1\sigma$. Black ellipses indicate structures of expanding molecular gas around the boundary of \snr. Three red arrows in the right panel mark the locations of three examples of broadened CO lines associated with \snr.
}
\label{f:f1}
\end{figure*}

We show the pseudo tricolor image of integrated intensities of three CO isotope lines of \snr\ over a wide velocity range in Figure~\ref{f:f1}. Multiple shell structures are clearly seen around a circular region with a radius of about $0.28^\circ$, which are distributed from the north to the southwest and in the east. \snr\ overlaps with a known SNR G16.0$-$0.5 but it is much larger in size. The molecular shell associated with G16.0$-$0.5 in its southern region is also visible, which was detected in previous work too \citep{Beaumont+2011}.
As shown in the position-velocity maps along horizontal and vertical strips across \snr\ (Figure~\ref{f:f1}), the molecular shells are expanding at a velocity of about 10 \km\ps.
Some molecular clumps correlated with the expanding shells are also strongly disturbed, indicated by line broadening features of about 20 \km\ps\ (see, for example, those indicated by three red arrows in the right panel of Figure~\ref{f:f1}). 
We note that the line broadening features are significant and not due to scanning effects (artifacts caused by baseline variations).
Two of these molecular clumps are located within and on the edge of the G16.0$-$0.5 region. However, the blueshifted velocity of these molecular clumps is consistent with the blueshifted velocity direction of the expanding shell at their locations.
The quiescent molecular gas at $\sim$+43 \km\ps\ around the expanding shell structures is rarefied.
Molecular gases at $\sim$+57 \km\ps\ inside \snr\ are all disturbed or associated with the expanding shells.
There is a small amount of quiescent molecular gas at $\sim$+57 \km\ps\ outside \snr.
However, this small amount of molecular gas may have a peculiar velocity compared to the Galactic rotation curve, but the major $\sim$+43 \km\ps\ component is unlikely.
The expanding shell structures and broadened line emission are probably associated with the quiescent molecular gas at $\sim$+43 \km\ps.
The $\sim$+43 and $\sim$+57 \km\ps\ MCs may be adjacent to each other and at the same distance, however, it is uncertain.

\begin{figure*}[ptbh!]
\centerline{{\hfil\hfil
\psfig{figure=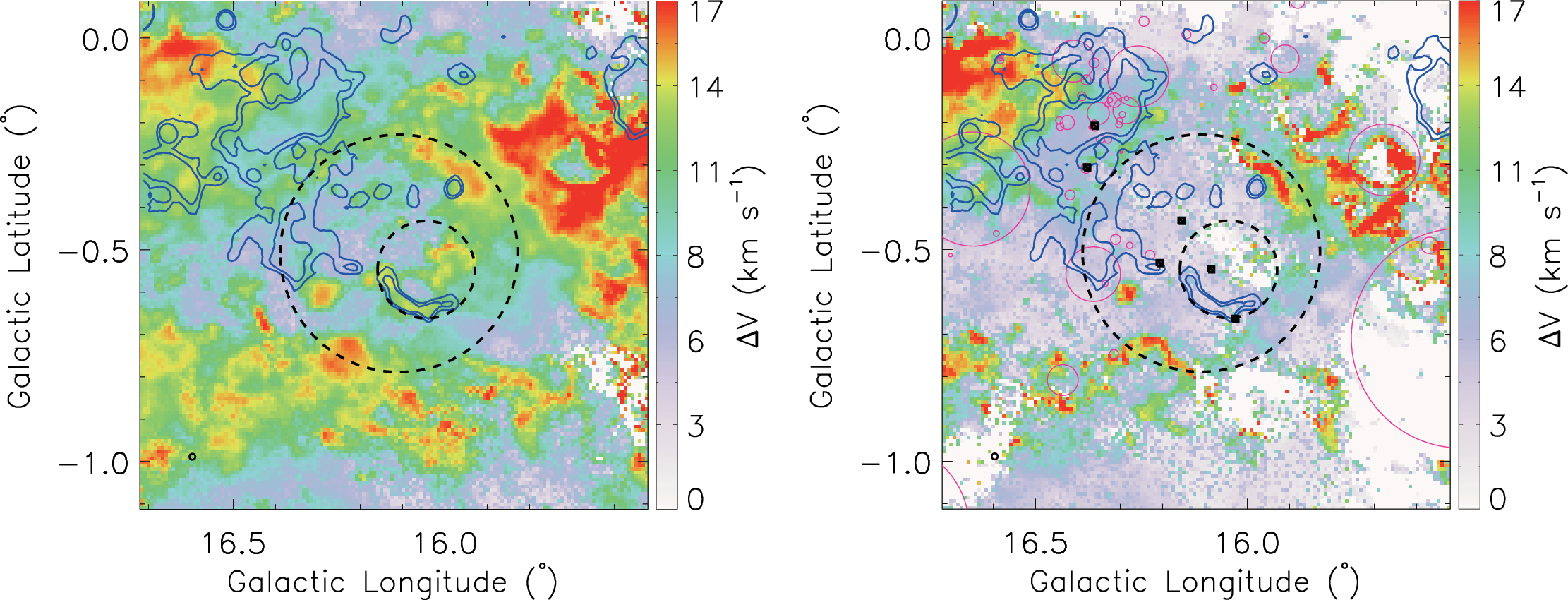,height=3in,angle=0, clip=}
\hfil\hfil}}
\caption{ Velocity dispersion (moment 2) maps of \twCO\ ({\it left}) and \thCO\ ({\it right}) line emission over the velocity range from +40 to +60 \km\ps, overlaid with the GLEAM 170--230 MHz radio-continuum emission contours ($\sim$2$'.25$ angular resolution). The contour levels are 0.76 and 0.81 Jy/beam. The black dashed circles are the same as in Figure~\ref{f:f1}, indicating the extent of \snr\ and SNR G16.0$-$0.5, respectively.
The \HII\ regions are indicated by magenta circles in the right panel, and all are introduced from the WISE catalog of Galactic \HII\ regions \citep{Anderson+2014}, which is one of the most complete catalogs of \HII\ regions in the Galaxy.
Radio continuum sources extracted from the THOR survey are marked by black solid boxes \citep{Bihr+2016, Wang+2018}.
}
\label{f:mom2}
\end{figure*}

The kinematic properties of molecular gases and the corresponding spatial distribution can be illustrated by the velocity dispersion map.
As shown in the \twCO\ velocity dispersion map (left panel, Figure~\ref{f:mom2}), disturbed molecular gases are present in the northwestern shell of \snr\ and in a molecular clump located around the southeastern boundary. There is also some diffuse and disturbed molecular gas distributed around the northeastern boundary of \snr, where there is a weak radio-continuum shell.
Known \HII\ regions are introduced from the WISE catalog of Galactic \HII\ regions \citep{Anderson+2014}, which is one of the most complete catalogs of \HII\ regions in the Galaxy (see the right panel of Figure~\ref{f:mom2}). Note that the systemic velocity of some \HII\ regions is not within the selected velocity range.
The disturbed molecular gases along the \snr\ boundary do not originate from known \HII\ regions.
For SNR G16.0$-$0.5, disturbed gas in the associated molecular shell is well correlated with the southeastern radio-continuum shell of the remnant.
The \thCO\ line emission is optically thinner than the \twCO\ line emission, which can trace inner and denser molecular gas. The \thCO\ velocity dispersion map (right panel, Figure~\ref{f:mom2}) shows the distribution of dense disturbed molecular gas. By comparison, no dense disturbed gas associated with G16.0$-$0.5 is seen. For \snr, dense disturbed gas distributed in a thin layer is seen inside its northwestern molecular shell.
Besides this, in its southeastern clump, dense disturbed gas is distributed only in the thin inner layer facing the center of \snr.
It indicates that the disturbed gas in the southeastern clump is mainly distributed on its surface and transmitted toward the outside of \snr.
Further study of these disturbed molecular gases in future work is needed for a better understanding of them.
For MCs outside the \snr\ region, some disturbed gases are associated with \HII\ regions, such as those in the northwest, northeast, and southeast.

\begin{figure*}[ptbh!]
\centerline{{\hfil\hfil
\psfig{figure=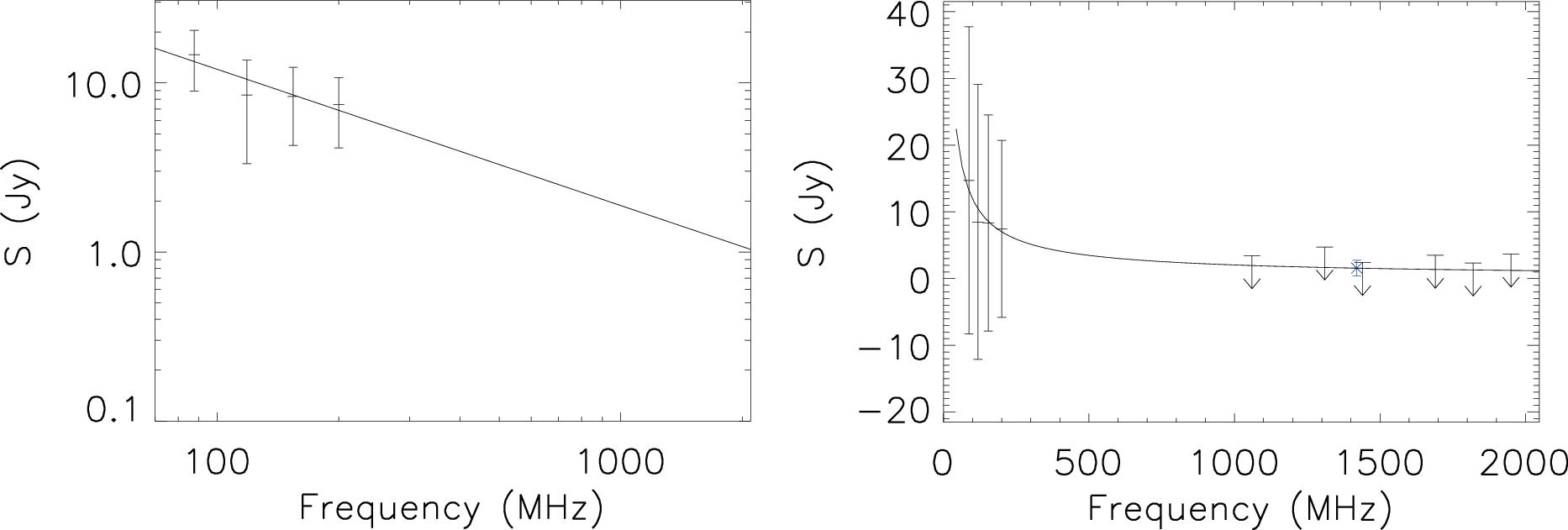,height=2.5in,angle=0, clip=}
\hfil\hfil}}
\caption{
Flux densities of \snr\ obtained from the GLEAM data only ({\it left}) and the GLEAM+THOR+(THOR+VGPS) data ({\it right}).
The flux density obtained from the THOR+VGPS 1.4 GHz continuum data is presented by a blue star in the right panel.
The flux densities are extracted from the \snr\ region indicated in Figure~\ref{f:mom2}. An annular region half the beamsize of the GLEAM 72--103 MHz data away from the source region, with an area similar to the source region, is used to subtract the background emission.
The flux densities of \HII\ regions, THOR radio-continuum sources, and the SNR G16.0$-$0.5 region, all enlarged by half the beamsize, were removed from the calculation (see regions in the right panel of Figure~\ref{f:mom2}).
The axes in the left panel are logarithmic, while the ones in the right panel are linear.
Internal and external flux density scale errors are applied as $2\%$ and $8\%$ for the GLEAM data in {\it left} and {\it right} panels, respectively \citep{Hurley-Walker+2017}. The flux density uncertainty of the THOR and THOR+VGPS data is determined by measuring the variation of the emission free region \citep[see][for reference]{Anderson+2017}.
The fit power-law spectra are shown by solid lines. The best-fit spectral indices are $-0.8\pm0.3$ and $-0.77\pm0.04$ for the GLEAM and GLEAM+(THOR+VGPS) data, respectively.
The upper limits of the flux densities from the THOR-only data are shown for comparison but are not used in fitting.
}
\label{f:radiospec}
\end{figure*}

\snr\ presents a partial radio-continuum shell along its northeastern boundary (see Figure~\ref{f:mom2}).
There is significant radio-continuum emission outside \snr\ in the northeast, which is associated with \HII\ regions (see the right panel of Figure~\ref{f:mom2}). However, radio-continuum emissions from \snr\ and the \HII\ regions surround different centers.
We measured the flux densities of \snr\ in different bands, with the background level subtracted. 
We first smoothed all the radio-continuum data in different bands to a common resolution as the lowest resolution GLEAM 72--103 MHz data.
An annular region half the beamsize of the GLEAM 72--103 MHz data away from the source region, with an area similar to the source region, was used to subtract the background emission. The flux densities of \HII\ regions, THOR radio-continuum sources, and the SNR G16.0$-$0.5 region, all enlarged by half the beamsize, are removed from the calculation (see regions in the right panel of Figure~\ref{f:mom2}).
The same regions are applied for all GLEAM, THOR, and THOR+VGPS bands. 
The radio-continuum emission of \snr\ is weak, hence, the individual measured flux densities are not so significant.
The overall radio-continuum spectrum can be fit by the power-law function (see Figure~\ref{f:radiospec}).
The best-fit spectral indices obtained using the GLEAM-only and GLEAM+(THOR+VGPS) data are consistent and steep, i.e.\ $\alpha=-0.8\pm0.3$ ($S_{\nu} \propto \nu ^{\alpha}$).
No radio-continuum emission is detected by the THOR-only data, which is likely to suffer from serious missing flux problems in the interferometric observations for such a large source.
These indicate that the radio-continuum emission is nonthermal and not from \HII\ regions.
The extrapolated 1~GHz surface brightness of \snr\ is estimated as $\Sigma_{1~GHz}\sim3\E{-21}$~\brightnessunit.

\section{Discussion}\label{sec:discuss}
\begin{figure*}[ptbh!]
\centerline{{\hfil\hfil
\psfig{figure=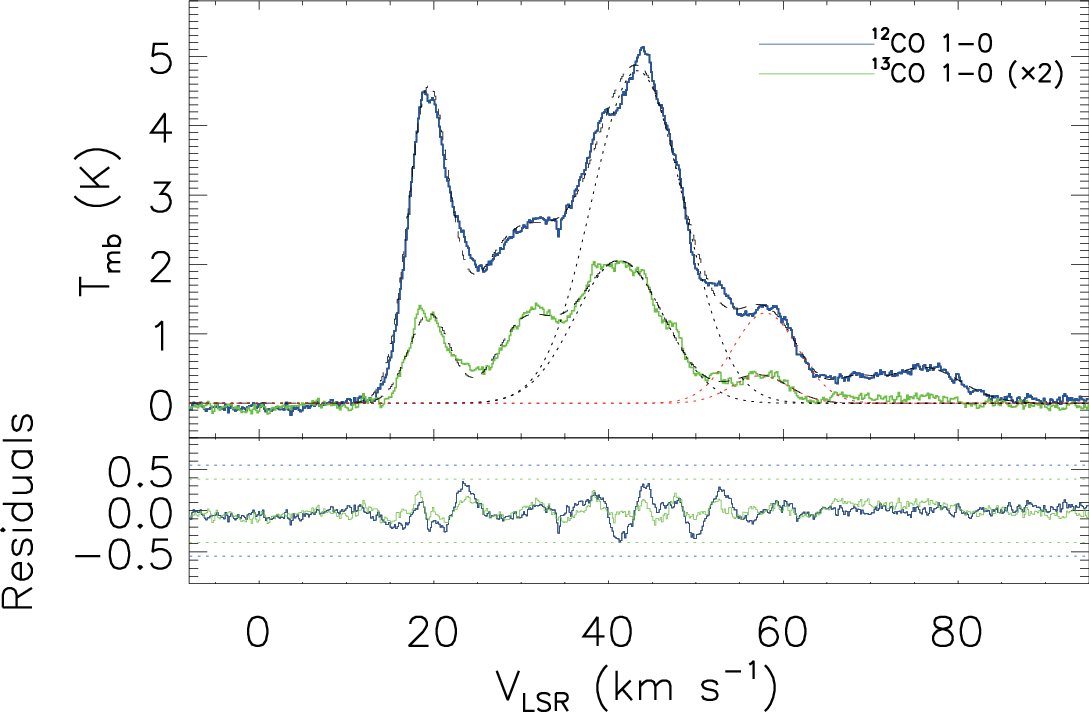,height=3.5in,angle=0, clip=}
\hfil\hfil}}
\caption{\twCO\ (J=1--0) (blue) and \thCO\ (J=1--0) (green) spectra extracted from 1.2 times enlarged \snr\ region with 1.2 times enlarged SNR G16.0$-$0.5 region subtracted, together with their best-fit Gaussian model and residuals.
The \twCO\ components peak at +19.3, +29.5, +43.3, +58.0, +68.1, and +76.0 \km\ps, and the \thCO\ components peak at +19.5, +30.2, +41.3, and +57.4 \km\ps.
Individual Gaussian models of the $\sim$+43 and $\sim$+57 \km\ps\ components are shown by black and red dotted lines, respectively.
$5\sigma$ levels of the residuals are shown by blue and green dotted lines for \twCO\ and \thCO, respectively. The \thCO\ (J=1--0) spectrum and its fitting result and residuals are multiplied by a factor of 2 for better visibility.
}
\label{f:cospec}
\end{figure*}

\begin{table*}
\small
\begin{center}
\caption{Fit and derived parameters for two related velocity components in \snr.\label{tab:para}}
\begin{tabular}{cccccc}
\hline\hline
\multicolumn{5}{c}{\sl Gaussian Components}\\
 Component & Line & Peak \tmb & Center \vlsr \tablefootmark{a} & FWHM \\
 & & (K)& (\km\ps) & (km~s$^{-1}$) \\
 \hline
rest&\twCO\ (J=1--0) &$4.79\pm0.03$ &$+43.31\pm0.07$ &$12.4\pm0.2$\\
&\thCO\ (J=1--0) &$1.027\pm0.007$ &$+41.25\pm0.09$ &$12.0\pm0.3$ \\
shifted&\twCO\ (J=1--0) &$1.30\pm0.03$ &$+58.0\pm0.2$ &$8.8\pm0.4$\\
&\thCO\ (J=1--0) & $0.198\pm0.008$ &$+57.4\pm0.2$ &$7.8\pm0.5$ \\
 \hline\hline
\multicolumn{6}{c}{\sl Derived Physical Parameters}\\
 Component & \tex\tablefootmark{b} & $\tau(^{13}$CO)\tablefootmark{b} & \ncol\tablefootmark{c} & $M$\tablefootmark{c} &\\
 & (K) & & $(10^{21}~\cm^{-2})$ & $(10^{4}~\Msun)$ &\\
\hline
rest& $8.1$ & $0.27$ & $11$ (11) & $14$\du$^2$ ($17$\du$^2$)&\\
shifted& $4.3$ & $0.19$ & $1.9$ (2.2) & $2.5$\du$^2$ ($3.2$\du$^2$)&\\
\hline
\end{tabular}
\tablefoot{
\tablefoottext{a}{\vlsr\ is the velocity with respect to the local standard of rest.}\\
\tablefoottext{b}{Using the assumption of local thermal equilibrium (LTE).
See the details of the calculation method in \cite{Zhou+2016}.}\\
\tablefoottext{c}{Derived from \thCO\ column density by assuming the \thCO\ abundance of 1.4$\E{-6}$ \citep{Ripple+2013}. For comparison, we also show the values in brackets, which are estimated by using the conversion factor $N({\rm H}_2)/{W}(^{12}{\rm CO)}\simeq1.8\times10^{20}~{\rm cm}^{-2}~{\rm K}^{-1}~{\rm km}^{-1}$~s \citep{Dame+2001}. \du\ stands for $d/(3.2~{\rm kpc})$, where d is the distance to \snr\ in units of kpc.}}
\end{center}
\end{table*}

We extracted CO spectra in the \snr\ region, which is defined as a circular region 1.2 times larger than that shown in Figure~\ref{f:mom2}.
We also excluded the SNR G16.0$-$0.5 region, which is a circular region 1.2 times larger than that shown in Figure~\ref{f:mom2}. The extracted \twCO\ and \thCO\ spectra are shown in Figure~\ref{f:cospec}. According to the number of their peaks, the \twCO\ spectrum is fit by six Gaussian functions, and the \thCO\ spectrum is fit by four Gaussian functions.
As shown in Section~\ref{sec:result}, \snr\ is associated with the MC at a systemic velocity of +41.3 \km\ps. Molecular gas at +57.4 \km\ps\ within the \snr\ region is mainly driven by \snr\ from the +41.3 \km\ps\ MC. The different velocities between the expanding shell of \snr\ and its parent MC indicate that the medium surrounding \snr\ has a density gradient in the line of sight.
The schematic diagram of the formation of the expanding shell of \snr\ is shown in the appendix. 
The fit and derived parameters for the +41.3 and +57.4 \km\ps\ components are listed in Table~\ref{tab:para}.
The kinematic distances of the +41.3 \km\ps\ MC are estimated as $3.2\pm0.4$ and $4.0\pm0.3$ kpc, based on a full distance probability density function\footnote{http://bessel.vlbi-astrometry.org/node/378} \citep{Reid+2016, Reid+2019}. The +41.3 \km\ps\ MC is more likely to be at 3.2 kpc, which has a probability of 0.53.
The momentum and kinetic energy of the expanding molecular gas of \snr\ are estimated to be $4\E{5}$\du$^2~\Msun\km\s^{-1}$ and $6.4\E{49}$\du$^2$ erg, respectively.
We note that, although a far kinematic distance of the +41.3 \km\ps\ MC can be estimated to be $\sim$12 kpc, the corresponding kinetic energy of the molecular gas would be too large, i.e.\ $9\E{50}$ erg, even comparable to the typical SN explosion energy value.
If the expanding shell is produced by the stellar wind of the OB star, even if all the stellar wind energy is transferred to the shell, a star with a mass greater than that of an O9 V type star is required \citep[e.g.,][]{Abbott1982}. However, we searched the SIMBAD astronomical database \citep{Wenger+2000} within a 1.2 times enlarged region of \snr\ and found no O-type stars. Only a few B-type stars are found in the region, which are located at distances of less than $\sim$1.2 kpc, and, hence, they are not related to \snr.
The expanding shell with thin and dense layers, and its physical properties, indicates an SN origin for \snr, which is further supported by the associated nonthermal radio-continuum emission.

If \snr\ is at a distance of 3.2 kpc, its radius is 15.6 pc.
The weak radio-continuum emission as well as the presence of dense shell structures indicate that SNR \snr\ is in the radiative stage.
Since the molecular shells are basically distributed on a circle on the plane of the sky, some simple models of the evolution of SNRs in a homogeneous medium can be applied as approximations to provide some references.
Assuming that the SNR evolved in a homogeneous ISM, its age can be estimated as $t=2r_s/(7v_s)\sim2.7\E{5}$\du~yr \citep{McKeeOstriker1977}, and the explosion energy as $E_{SN}=6.8\E{43} n_0^{1.16} (v_s/1~{\rm km~s}^{-1})^{1.35} (r_s/1~{\rm pc})^{3.16} \zeta_m^{0.161}\sim2\E{51}$\du$^2$~erg, where the ambient hydrogen density $n_0$ is estimated to be $\sim$63 cm$^{-3}$, the shock velocity $v_s$ as $v_{mol}=16.1$ \km\ps, the radius of the SNR $r_s$ 15.6 pc, and the metallicity parameter $\zeta_m=Z/Z_{\rm \odot}$ is set to be 1 \citep{Cioffi+1988}.
The ambient density is estimated by assuming that the mass of the expanding molecular shell was initially uniformly distributed throughout its internal volume. 
The explosion energy obtained is big in comparison to the typical value, i.e.\ $10^{51}$ erg.
As the SNR evolves in a non-uniform medium, there is a leakage of the SNR energy in less dense directions, and therefore the explosion energy is even underestimated.
Considering that the SNR is likely to have evolved in a bubble blown by the stellar wind of its progenitor, with the contribution of the stellar wind to the accumulation of the ISM, less SN explosion energy will be required.
In the wind-blown bubble scenario, we estimate the explosion energy and age using the model of \cite{Chen+2003}.
Assuming $\lambda=r_j/r_s=0.9$ with $(\eta, \beta)=(1, 0)$, its explosion energy can be estimated as $E_{SN}=1.05 \rho_0 v_s^2 r_s^3 \lambda^{-2} [F_v^{(R)}(\lambda)]^{-2}\sim10^{51}$\du$^2$ erg and the age as $t=1.02 \rho_0^{0.5} r_s^{2.5} E_{SN}^{-0.5} \lambda^{-1} F_r^{(R)}(\lambda)\sim10^{5}$\du~yr, where $r_j$ is the radius of the progenitor's wind-blown bubble, $\rho_0$ is the density of the wall of the bubble, the dimensionless term $F_v^{(R)}(\lambda)$ is estimated to be $\sim$0.444, and $F_r^{(R)}(\lambda)$ is estimated to be $\sim$0.206 \citep[cf.\ Equations [26{]}, [29{]}, [31{]}, and [35{]} in][]{Chen+2003}.
For $\lambda=r_j/r_s=0.9$, the radius of the wind-blown bubble is $r_j\sim14$ pc.
The assumption of $\eta=1$ means that the SNR enters the radiative phase directly after hitting the cavity wall, which is reasonable for $r_j\gtrsim3.8$~pc \citep[see Section 2.1 in][]{Chen+2003}, and the assumption of $\beta=0$ means that the density inside the cavity is much smaller than that of the cavity wall.
The density of the wall of the wind-blown bubble is estimated by assuming that the mass of the expanding molecular shell was initially evenly distributed in a spherical shell with inner and outer radii of $r_j$ and $r_s$, respectively, i.e.\ $n_0$$\sim$233 cm$^{-3}$.
According to the radius of the wind-blown bubble, the progenitor's initial mass can be estimated to be $\sim$19~$\Msun$ by applying the linear relationship between the size of a massive star's main-sequence bubble in a molecular environment and its initial mass \citep{Chen+2013}.
In the wind-blown bubble scenario, a relatively reasonable explosion energy can be obtained. Note that the uniform wind-blown bubble model is still a rough approximation.
Since the radius of the SNR on the plane of the sky is smaller than the average radius (see the appendix for more information), the age obtained (i.e., $\sim$$10^5$~yr) is only a lower limit.
Further analysis in future work is needed to obtain more appropriate models based on specific molecular gas distributions (e.g., numerical models).

The radio-continuum emission from \snr\ is weak, of which the diameter ($\sim$31 pc) and the radio-continuum surface brightness ($\Sigma_{1~GHz}\sim3\E{-21}$~\brightnessunit) conform to the $\Sigma$--D relation \citep{Pavlovic+2018}.
In general, as the SNRs become older, their radio-continuum emissions become weaker, e.g., the SNRs discovered by \HI\ observations with an age of around one million years is with $\Sigma_{1 GHz}\lesssim7\E{-23}$~\brightnessunit \citep{Koo+2006, XiaoZhu2014}.
Nevertheless, when the SNR shock encounters dense clouds, due to the compressed magnetic fields and the accelerated preexisting cosmic ray electron, the radio-continuum emission can be enhanced \citep{BlandfordCowie1982, DraineMcKee1993}. It seems to be efficient for SNRs encountering clouds while still having quite a high Mach number \citep{Pavlovic+2018}. No radio-continuum emission is detected around the northwestern and southeastern molecular shells, which are probably swept up by the SNR shock but not preexisting ones.
The nonthermal spectral index of the radio-continuum emission from \snr\ is steep, which resembles indices of shell-type SNRs detected in the GLEAM survey \citep[e.g.,][]{Hurley-Walker+2019c}.
The steep spectral index can be expected for an old SNR due to the energy loss of electrons through synchrotron radiation \citep[e.g.,][]{XiaoZhu2014}.

\snr\ completely covers SNR G16.0$-$0.5 and is larger in size.
\cite{Beaumont+2011} used a machine learning algorithm to classify broadened \twCO\ (J=3--2) line structures in a position-position-velocity data cube for SNR G16.0$-$0.5.
Most of the coverage of their data is in the \snr\ region.
In addition to most of the identified broad lines distributed within SNR G16.0$-$0.5, there are also broad lines identified outside G16.0$-$0.5 but within \snr\ \citep[see, for example, Figure 8 in][]{Beaumont+2011}.
As shown in Section~\ref{sec:result} (Figure~\ref{f:f1}) here, there are broad lines both inside and outside the G16.0$-$0.5 region correlated with the expanding CO shell of \snr.
Even if it is correlated with \snr, we cannot totally rule out the possibility that the broad lines in the G16.0$-$0.5 region are associated with SNR G16.0$-$0.5. For old SNRs, their shock waves are weak, and it is difficult to accelerate the molecular gas to very high velocities. The molecular clumps with broad line features, which are correlated with \snr, may have been accelerated earlier by the fast SNR shock and they may have experienced less deceleration due to the greater density.
The molecular gas associated with SNR G16.0$-$0.5 is mainly at $\sim$+68 \km\ps, and the $\sim$+68 \km\ps\ velocity component is an independent MC distributed both inside and outside SNR G16.0$-$0.5 \citep[see][]{Zhou+2023}.
Accordingly, most of the broadened CO lines associated with SNR G16.0$-$0.5 are blueshifted.
If the molecular gas at $\sim$+68 \km\ps\ within SNR G16.0$-$0.5 is redshifted from the +41.3 \km\ps\ component by the stellar wind of its progenitor, it should be further redshifted by the shock wave of the subsequent remnant, which is not the case.
It is still possible that SNR G16.0$-$0.5 is located near the shell on the farside of \snr, but there is no evidence to support this.

In conclusion, we suggest \snr\ is an old SNR.
As a background to SNR G16.0$-$0.5, some broad lines in the G16.0$-$0.5 region show correlations with \snr.
The finding of such an SNR characterized by an expanding CO shell may suggest other similar remnants waiting to be discovered.

\begin{acknowledgements}
We thank Ping Zhou for helpful feedback on the draft manuscript which improved the paper.
We also thank the anonymous referee for very valuable comments that improved this paper and its conclusions.
We are grateful to all the members of the Milky Way Scroll Painting-CO line survey group, especially the staff of the Qinghai Radio Observing Station at Delingha, for their support during the observation.
In particular, we thank the anonymous referee of \cite{Zhou+2023} for providing very helpful suggestions that inspired our in-depth study of this region.
This work is part of the Milky Way Imaging Scroll Painting (MWISP) multi-line survey project, which is supported by the National Key R\&D Program of China grant no.\ 2023YFA1608000 and 2017YFA0402701, and the Key Research Program of Frontier Sciences, CAS, grant no.\ QYZDJ-SSW-SLH047.
Y.S., J.Y., and Y.C. acknowledge support from the NSFC grants 12173090, 12041305, 12173018 and 12121003.
This research has made use of the SIMBAD database operated at CDS, Strasbourg, France.
We acknowledge the use of the VGPS data; the National Radio Astronomy Observatory is a facility of the National Science Foundation operated under cooperative agreement by Associated Universities, Inc.
\end{acknowledgements}

\paragraph{Data Availability}
The MWISP CO~(J=1--0) data of SNR \snr\ is available online in https://www.scidb.cn/en, at https://doi.org/10.57760/sciencedb.15577.

\bibliographystyle{aa} 

\begin{thebibliography}{52}
\expandafter\ifx\csname natexlab\endcsname\relax\def\natexlab#1{#1}\fi

\bibitem[{{Abbott}(1982)}]{Abbott1982}
{Abbott}, D.~C. 1982, \apj, 263, 723

\bibitem[{{Ambrocio-Cruz} {et~al.}(2006){Ambrocio-Cruz}, {Rosado}, \& {de La
  Fuente}}]{Ambrocio-Cruz+2006}
{Ambrocio-Cruz}, P., {Rosado}, M., \& {de La Fuente}, E. 2006, \rmxaa, 42, 241

\bibitem[{{Anderson} {et~al.}(2014){Anderson}, {Bania}, {Balser}, {Cunningham},
  {Wenger}, {Johnstone}, \& {Armentrout}}]{Anderson+2014}
{Anderson}, L.~D., {Bania}, T.~M., {Balser}, D.~S., {et~al.} 2014, \apjs, 212,
  1

\bibitem[{{Anderson} {et~al.}(2017){Anderson}, {Wang}, {Bihr}, {Rugel},
  {Beuther}, {Bigiel}, {Churchwell}, {Glover}, {Goodman}, {Henning}, {Heyer},
  {Klessen}, {Linz}, {Longmore}, {Menten}, {Ott}, {Roy}, {Soler}, {Stil}, \&
  {Urquhart}}]{Anderson+2017}
{Anderson}, L.~D., {Wang}, Y., {Bihr}, S., {et~al.} 2017, \aap, 605, A58

\bibitem[{{Ball} {et~al.}(2023){Ball}, {Kothes}, {Rosolowsky}, {West},
  {Becker}, {Filipovi{\'c}}, {Gaensler}, {Hopkins}, {Koribalski}, {Landecker},
  {Leahy}, {Marvil}, {Sun}, {Bufano}, {Carretti}, {Ingallinera}, {Van Eck}, \&
  {Willis}}]{Ball+2023}
{Ball}, B.~D., {Kothes}, R., {Rosolowsky}, E., {et~al.} 2023, \mnras, 524, 1396

\bibitem[{{Beaumont} {et~al.}(2011){Beaumont}, {Williams}, \&
  {Goodman}}]{Beaumont+2011}
{Beaumont}, C.~N., {Williams}, J.~P., \& {Goodman}, A.~A. 2011, \apj, 741, 14

\bibitem[{{Becker} {et~al.}(2021){Becker}, {Hurley-Walker}, {Weinberger},
  {Nicastro}, {Mayer}, {Merloni}, \& {Sanders}}]{Becker+2021}
{Becker}, W., {Hurley-Walker}, N., {Weinberger}, C., {et~al.} 2021, \aap, 648,
  A30

\bibitem[{{Beuther} {et~al.}(2016){Beuther}, {Bihr}, {Rugel}, {Johnston},
  {Wang}, {Walter}, {Brunthaler}, {Walsh}, {Ott}, {Stil}, {Henning},
  {Schierhuber}, {Kainulainen}, {Heyer}, {Goldsmith}, {Anderson}, {Longmore},
  {Klessen}, {Glover}, {Urquhart}, {Plume}, {Ragan}, {Schneider},
  {McClure-Griffiths}, {Menten}, {Smith}, {Roy}, {Shanahan}, {Nguyen-Luong}, \&
  {Bigiel}}]{Beuther+2016}
{Beuther}, H., {Bihr}, S., {Rugel}, M., {et~al.} 2016, \aap, 595, A32

\bibitem[{{Bihr} {et~al.}(2016){Bihr}, {Johnston}, {Beuther}, {Anderson},
  {Ott}, {Rugel}, {Bigiel}, {Brunthaler}, {Glover}, {Henning}, {Heyer},
  {Klessen}, {Linz}, {Longmore}, {McClure-Griffiths}, {Menten}, {Plume},
  {Schierhuber}, {Shanahan}, {Stil}, {Urquhart}, \& {Walsh}}]{Bihr+2016}
{Bihr}, S., {Johnston}, K.~G., {Beuther}, H., {et~al.} 2016, \aap, 588, A97

\bibitem[{{Blandford} \& {Cowie}(1982)}]{BlandfordCowie1982}
{Blandford}, R.~D. \& {Cowie}, L.~L. 1982, \apj, 260, 625

\bibitem[{{Chen} {et~al.}(2014){Chen}, {Jiang}, {Zhou}, {Su}, {Zhou}, {Li}, \&
  {Zhang}}]{Chen+2014}
{Chen}, Y., {Jiang}, B., {Zhou}, P., {et~al.} 2014, in IAU Symposium, Vol. 296,
  Supernova Environmental Impacts, ed. A.~{Ray} \& R.~A. {McCray}, 170--177

\bibitem[{{Chen} {et~al.}(2003){Chen}, {Zhang}, {Williams}, \&
  {Wang}}]{Chen+2003}
{Chen}, Y., {Zhang}, F., {Williams}, R.~M., \& {Wang}, Q.~D. 2003, \apj, 595,
  227

\bibitem[{{Chen} {et~al.}(2013){Chen}, {Zhou}, \& {Chu}}]{Chen+2013}
{Chen}, Y., {Zhou}, P., \& {Chu}, Y.-H. 2013, \apjl, 769, L16

\bibitem[{{Churazov} {et~al.}(2021){Churazov}, {Khabibullin}, {Bykov},
  {Chugai}, {Sunyaev}, \& {Zinchenko}}]{Churazov+2021}
{Churazov}, E.~M., {Khabibullin}, I.~I., {Bykov}, A.~M., {et~al.} 2021, \mnras,
  507, 971

\bibitem[{{Cioffi} {et~al.}(1988){Cioffi}, {McKee}, \&
  {Bertschinger}}]{Cioffi+1988}
{Cioffi}, D.~F., {McKee}, C.~F., \& {Bertschinger}, E. 1988, \apj, 334, 252

\bibitem[{{Dame} {et~al.}(2001){Dame}, {Hartmann}, \& {Thaddeus}}]{Dame+2001}
{Dame}, T.~M., {Hartmann}, D., \& {Thaddeus}, P. 2001, \apj, 547, 792

\bibitem[{{Dokara} {et~al.}(2021){Dokara}, {Brunthaler}, {Menten}, {Dzib},
  {Reich}, {Cotton}, {Anderson}, {Chen}, {Gong}, {Medina}, {Ortiz-Le{\'o}n},
  {Rugel}, {Urquhart}, {Wyrowski}, {Yang}, {Beuther}, {Billington}, {Csengeri},
  {Carrasco-Gonz{\'a}lez}, \& {Roy}}]{Dokara+2021}
{Dokara}, R., {Brunthaler}, A., {Menten}, K.~M., {et~al.} 2021, \aap, 651, A86

\bibitem[{{Draine} \& {McKee}(1993)}]{DraineMcKee1993}
{Draine}, B.~T. \& {McKee}, C.~F. 1993, \araa, 31, 373

\bibitem[{{Ferrand} \& {Safi-Harb}(2012)}]{FerrandSafi-Harb2012}
{Ferrand}, G. \& {Safi-Harb}, S. 2012, Advances in Space Research, 49, 1313

\bibitem[{{Fesen} {et~al.}(2020){Fesen}, {Weil}, {Raymond}, {Huet},
  {Rusterholz}, {di Cicco}, {Mittelman}, {Walker}, {Drechsler}, \&
  {Faworski}}]{Fesen+2020}
{Fesen}, R.~A., {Weil}, K.~E., {Raymond}, J.~C., {et~al.} 2020, \mnras, 498,
  5194

\bibitem[{{Gao} {et~al.}(2020){Gao}, {Reich}, {Reich}, {Hou}, \&
  {Han}}]{Gao+2020}
{Gao}, X.~Y., {Reich}, P., {Reich}, W., {Hou}, L.~G., \& {Han}, J.~L. 2020,
  \mnras, 493, 2188

\bibitem[{{Green}(2019)}]{Green2019}
{Green}, D.~A. 2019, Journal of Astrophysics and Astronomy, 40, 36

\bibitem[{{Hurley-Walker} {et~al.}(2017){Hurley-Walker}, {Callingham},
  {Hancock}, {Franzen}, {Hindson}, {Kapi{\'n}ska}, {Morgan}, {Offringa},
  {Wayth}, {Wu}, {Zheng}, {Murphy}, {Bell}, {Dwarakanath}, {For}, {Gaensler},
  {Johnston-Hollitt}, {Lenc}, {Procopio}, {Staveley-Smith}, {Ekers}, {Bowman},
  {Briggs}, {Cappallo}, {Deshpande}, {Greenhill}, {Hazelton}, {Kaplan},
  {Lonsdale}, {McWhirter}, {Mitchell}, {Morales}, {Morgan}, {Oberoi}, {Ord},
  {Prabu}, {Shankar}, {Srivani}, {Subrahmanyan}, {Tingay}, {Webster},
  {Williams}, \& {Williams}}]{Hurley-Walker+2017}
{Hurley-Walker}, N., {Callingham}, J.~R., {Hancock}, P.~J., {et~al.} 2017,
  \mnras, 464, 1146

\bibitem[{{Hurley-Walker} {et~al.}(2019{\natexlab{a}}){Hurley-Walker},
  {Filipovi{\'c}}, {Gaensler}, {Leahy}, {Hancock}, {Franzen}, {Offringa},
  {Callingham}, {Hindson}, {Wu}, {Bell}, {For}, {Johnston-Hollitt},
  {Kapi{\'n}ska}, {Morgan}, {Murphy}, {McKinley}, {Procopio}, {Staveley-Smith},
  {Wayth}, \& {Zheng}}]{Hurley-Walker+2019c}
{Hurley-Walker}, N., {Filipovi{\'c}}, M.~D., {Gaensler}, B.~M., {et~al.}
  2019{\natexlab{a}}, \pasa, 36, e045

\bibitem[{{Hurley-Walker} {et~al.}(2019{\natexlab{b}}){Hurley-Walker},
  {Hancock}, {Franzen}, {Callingham}, {Offringa}, {Hindson}, {Wu}, {Bell},
  {For}, {Gaensler}, {Johnston-Hollitt}, {Kapi{\'n}ska}, {Morgan}, {Murphy},
  {McKinley}, {Procopio}, {Staveley-Smith}, {Wayth}, \&
  {Zheng}}]{Hurley-Walker+2019a}
{Hurley-Walker}, N., {Hancock}, P.~J., {Franzen}, T.~M.~O., {et~al.}
  2019{\natexlab{b}}, \pasa, 36, e047

\bibitem[{{Kang} \& {Koo}(2007)}]{KangKoo2007}
{Kang}, J.-h. \& {Koo}, B.-C. 2007, \apjs, 173, 85

\bibitem[{{Kang} {et~al.}(2012){Kang}, {Koo}, \& {Salter}}]{Kang+2012}
{Kang}, J.-h., {Koo}, B.-C., \& {Salter}, C. 2012, \aj, 143, 75

\bibitem[{{Khabibullin} {et~al.}(2023){Khabibullin}, {Churazov}, {Bykov},
  {Chugai}, \& {Sunyaev}}]{Khabibullin+2023}
{Khabibullin}, I.~I., {Churazov}, E.~M., {Bykov}, A.~M., {Chugai}, N.~N., \&
  {Sunyaev}, R.~A. 2023, \mnras, 521, 5536

\bibitem[{{Koo} \& {Kang}(2004)}]{KooKang2004}
{Koo}, B.-C. \& {Kang}, J.-h. 2004, \mnras, 349, 983

\bibitem[{{Koo} {et~al.}(2006){Koo}, {Kang}, \& {Salter}}]{Koo+2006}
{Koo}, B.-C., {Kang}, J.-h., \& {Salter}, C.~J. 2006, \apjl, 643, L49

\bibitem[{{Koo} {et~al.}(2020){Koo}, {Kim}, {Park}, \& {Ostriker}}]{Koo+2020}
{Koo}, B.-C., {Kim}, C.-G., {Park}, S., \& {Ostriker}, E.~C. 2020, \apj, 905,
  35

\bibitem[{{Lu} {et~al.}(2021){Lu}, {Yan}, {Wen}, \& {Fang}}]{Lu+2021}
{Lu}, C.-Y., {Yan}, J.-W., {Wen}, L., \& {Fang}, J. 2021, Research in Astronomy
  and Astrophysics, 21, 033

\bibitem[{{McKee} \& {Ostriker}(1977)}]{McKeeOstriker1977}
{McKee}, C.~F. \& {Ostriker}, J.~P. 1977, \apj, 218, 148

\bibitem[{{Pavlovi{\'c}} {et~al.}(2018){Pavlovi{\'c}}, {Uro{\v{s}}evi{\'c}},
  {Arbutina}, {Orlando}, {Maxted}, \& {Filipovi{\'c}}}]{Pavlovic+2018}
{Pavlovi{\'c}}, M.~Z., {Uro{\v{s}}evi{\'c}}, D., {Arbutina}, B., {et~al.} 2018,
  \apj, 852, 84

\bibitem[{{Reich} \& {Sun}(2019)}]{ReichSun2019}
{Reich}, W. \& {Sun}, X.-H. 2019, Research in Astronomy and Astrophysics, 19,
  045

\bibitem[{{Reid} {et~al.}(2016){Reid}, {Dame}, {Menten}, \&
  {Brunthaler}}]{Reid+2016}
{Reid}, M.~J., {Dame}, T.~M., {Menten}, K.~M., \& {Brunthaler}, A. 2016, \apj,
  823, 77

\bibitem[{{Reid} {et~al.}(2019){Reid}, {Menten}, {Brunthaler}, {Zheng}, {Dame},
  {Xu}, {Li}, {Sakai}, {Wu}, {Immer}, {Zhang}, {Sanna}, {Moscadelli}, {Rygl},
  {Bartkiewicz}, {Hu}, {Quiroga-Nu{\~n}ez}, \& {van Langevelde}}]{Reid+2019}
{Reid}, M.~J., {Menten}, K.~M., {Brunthaler}, A., {et~al.} 2019, \apj, 885, 131

\bibitem[{Ripple {et~al.}(2013)Ripple, Heyer, Gutermuth, Snell, \&
  Brunt}]{Ripple+2013}
Ripple, F., Heyer, M.~H., Gutermuth, R., Snell, R.~L., \& Brunt, C.~M. 2013,
  \mnras, 431, 1296

\bibitem[{{Shan} {et~al.}(2012){Shan}, {Yang}, {Shi}, {Yao}, {Zuo}, {Lin},
  {Chen}, {Zhang}, {Duan}, {Cao}, {Li}, {Li}, {Liu}, \& {Zhong}}]{Shan+2012}
{Shan}, W.~L., {Yang}, J., {Shi}, S.~C., {et~al.} 2012, IEEE Transactions on
  Terahertz Science and Technology, 2, 593

\bibitem[{{Sofue}(2021)}]{Sofue2021}
{Sofue}, Y. 2021, Galaxies, 9, 13

\bibitem[{{Sofue} {et~al.}(2021){Sofue}, {Kohno}, \& {Umemoto}}]{Sofue+2021}
{Sofue}, Y., {Kohno}, M., \& {Umemoto}, T. 2021, \apjs, 253, 17

\bibitem[{{Stil} {et~al.}(2006){Stil}, {Taylor}, {Dickey}, {Kavars}, {Martin},
  {Rothwell}, {Boothroyd}, {Lockman}, \& {McClure-Griffiths}}]{Stil+2006}
{Stil}, J.~M., {Taylor}, A.~R., {Dickey}, J.~M., {et~al.} 2006, \aj, 132, 1158

\bibitem[{{Su} {et~al.}(2017){Su}, {Zhou}, {Yang}, {Chen}, {Chen}, {Gong}, \&
  {Zhang}}]{Su+2017b}
{Su}, Y., {Zhou}, X., {Yang}, J., {et~al.} 2017, \apj, 845, 48

\bibitem[{{Tammann} {et~al.}(1994){Tammann}, {Loeffler}, \&
  {Schroeder}}]{Tammann+1994}
{Tammann}, G.~A., {Loeffler}, W., \& {Schroeder}, A. 1994, \apjs, 92, 487

\bibitem[{{Toledo-Roy} {et~al.}(2014){Toledo-Roy}, {Vel{\'a}zquez}, {Esquivel},
  \& {Giacani}}]{Toledo-Roy+2014}
{Toledo-Roy}, J.~C., {Vel{\'a}zquez}, P.~F., {Esquivel}, A., \& {Giacani}, E.
  2014, \mnras, 437, 898

\bibitem[{{Wang} {et~al.}(2020){Wang}, {Beuther}, {Rugel}, {Soler}, {Stil},
  {Ott}, {Bihr}, {McClure-Griffiths}, {Anderson}, {Klessen}, {Goldsmith},
  {Roy}, {Glover}, {Urquhart}, {Heyer}, {Linz}, {Smith}, {Bigiel}, {Dempsey},
  \& {Henning}}]{Wang+2020}
{Wang}, Y., {Beuther}, H., {Rugel}, M.~R., {et~al.} 2020, \aap, 634, A83

\bibitem[{{Wang} {et~al.}(2018){Wang}, {Bihr}, {Rugel}, {Beuther}, {Johnston},
  {Ott}, {Soler}, {Brunthaler}, {Anderson}, {Urquhart}, {Klessen}, {Linz},
  {McClure-Griffiths}, {Glover}, {Menten}, {Bigiel}, {Hoare}, \&
  {Longmore}}]{Wang+2018}
{Wang}, Y., {Bihr}, S., {Rugel}, M., {et~al.} 2018, \aap, 619, A124

\bibitem[{{Wayth} {et~al.}(2015){Wayth}, {Lenc}, {Bell}, {Callingham},
  {Dwarakanath}, {Franzen}, {For}, {Gaensler}, {Hancock}, {Hindson},
  {Hurley-Walker}, {Jackson}, {Johnston-Hollitt}, {Kapi{\'n}ska}, {McKinley},
  {Morgan}, {Offringa}, {Procopio}, {Staveley-Smith}, {Wu}, {Zheng}, {Trott},
  {Bernardi}, {Bowman}, {Briggs}, {Cappallo}, {Corey}, {Deshpande}, {Emrich},
  {Goeke}, {Greenhill}, {Hazelton}, {Kaplan}, {Kasper}, {Kratzenberg},
  {Lonsdale}, {Lynch}, {McWhirter}, {Mitchell}, {Morales}, {Morgan}, {Oberoi},
  {Ord}, {Prabu}, {Rogers}, {Roshi}, {Shankar}, {Srivani}, {Subrahmanyan},
  {Tingay}, {Waterson}, {Webster}, {Whitney}, {Williams}, \&
  {Williams}}]{Wayth+2015}
{Wayth}, R.~B., {Lenc}, E., {Bell}, M.~E., {et~al.} 2015, \pasa, 32, e025

\bibitem[{{Wenger} {et~al.}(2000){Wenger}, {Ochsenbein}, {Egret}, {Dubois},
  {Bonnarel}, {Borde}, {Genova}, {Jasniewicz}, {Lalo{\"e}}, {Lesteven}, \&
  {Monier}}]{Wenger+2000}
{Wenger}, M., {Ochsenbein}, F., {Egret}, D., {et~al.} 2000, \aaps, 143, 9

\bibitem[{{Xiao} \& {Zhu}(2014)}]{XiaoZhu2014}
{Xiao}, L. \& {Zhu}, M. 2014, \mnras, 438, 1081

\bibitem[{{Zhou} {et~al.}(2023){Zhou}, {Su}, {Yang}, {Chen}, {Sun}, {Jiang},
  {Wang}, {Wang}, {Zhang}, {Xu}, {Yan}, {Yuan}, {Chen}, {Ao}, \&
  {Ma}}]{Zhou+2023}
{Zhou}, X., {Su}, Y., {Yang}, J., {et~al.} 2023, \apjs, 268, 61

\bibitem[{{Zhou} {et~al.}(2016){Zhou}, {Yang}, {Fang}, {Su}, {Sun}, \&
  {Chen}}]{Zhou+2016}
{Zhou}, X., {Yang}, J., {Fang}, M., {et~al.} 2016, \apj, 833, 4

\end{thebibliography}

\begin{appendix} 
\section{The schematic diagram}\label{sec:appendix}
\begin{figure*}[ptbh!]
\centerline{{\hfil\hfil
\psfig{figure=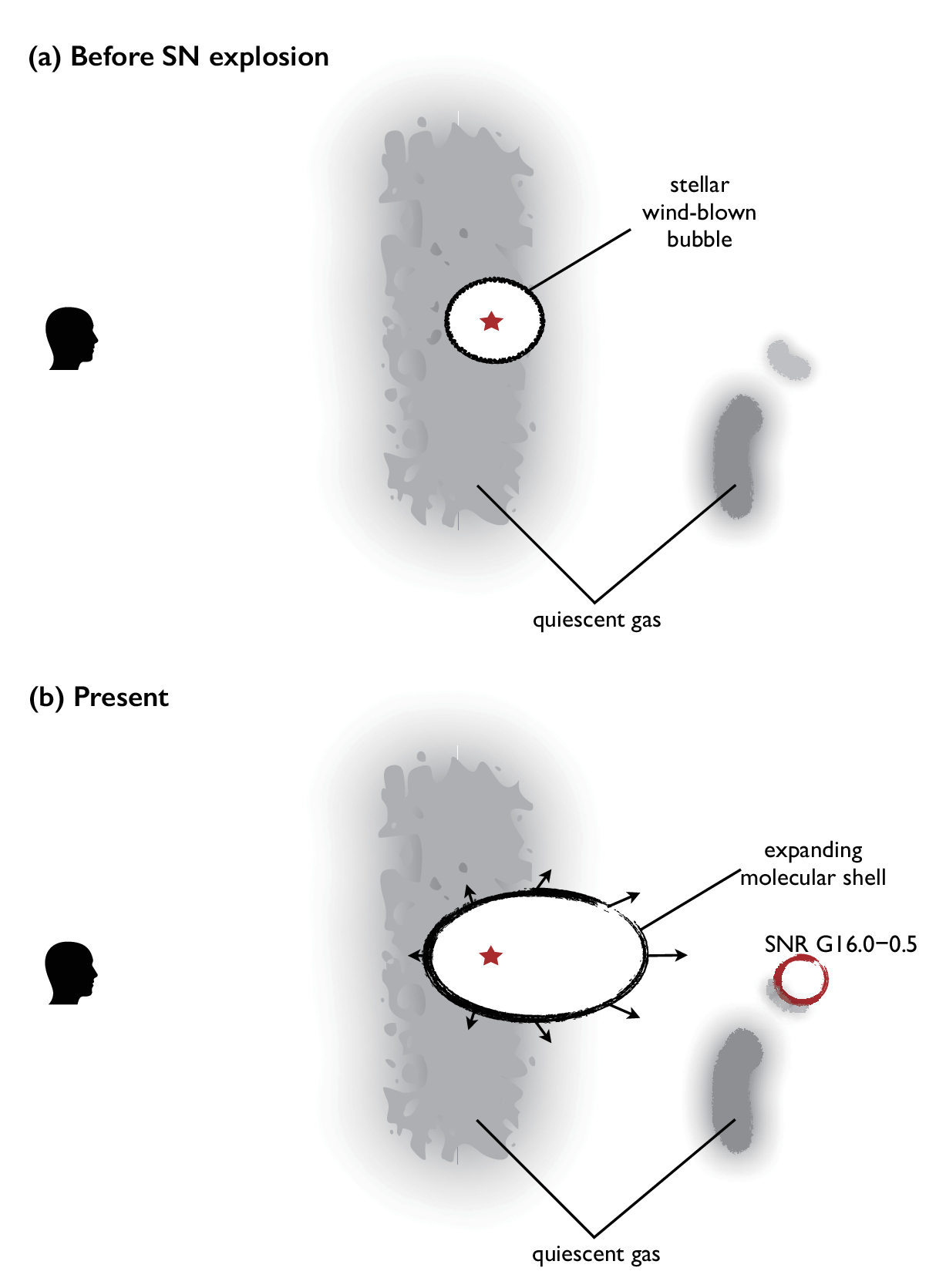,height=7in,angle=0, clip=}
\hfil\hfil}}
\caption{Schematic view of formation of expanding shell of \snr. 
The red star denotes the location of the initial SN explosion. The approximate location of SNR G16.0$-$0.5 is also marked with a red circle.
The arrows indicate the velocities in different parts of the shell. According to the distribution of the velocity component in the line of sight of different sections of the shell, a similar structure will be present in the position-velocity map as shown in Figure~\ref{f:f1}.
}
\label{f:shematicdiag}
\end{figure*}

\begin{figure*}[ptbh!]
\centerline{{\hfil\hfil
\psfig{figure=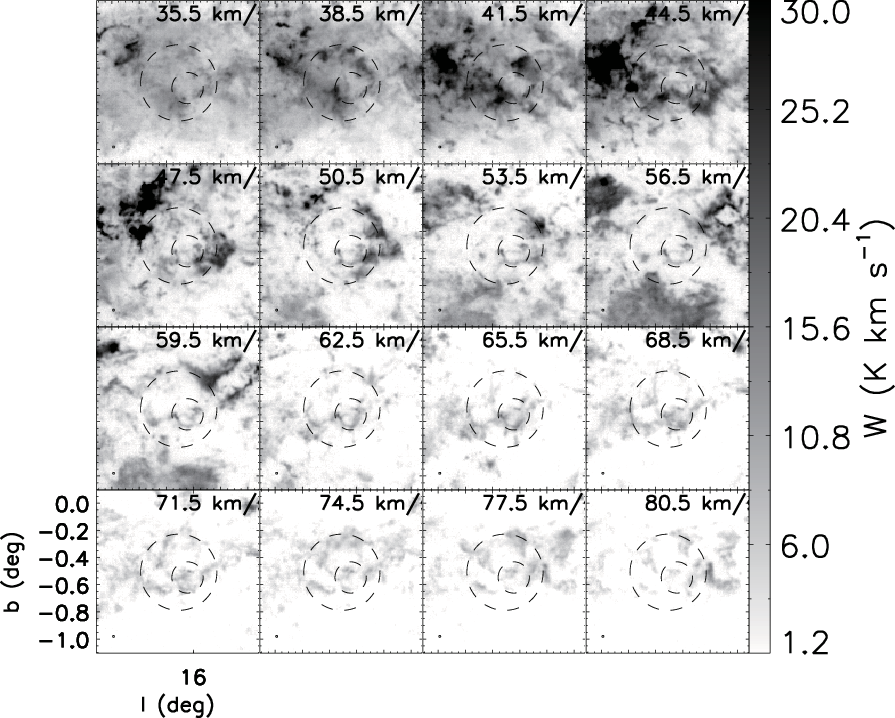,height=5in,angle=0, clip=}
\hfil\hfil}}
\caption{\twCO~(J=1--0) intensity maps integrated over each 3~\km\ps. The black dashed circles are the same as in Figure~\ref{f:f1}. Central velocities are indicated in each panel. The minimum value of each map is $3\sigma$.
}
\label{f:12stamp}
\end{figure*}

We present a schematic diagram in Figure~\ref{f:shematicdiag} to illustrate the formation of the expanding shell of \snr. As shown in the schematic diagram, when an SNR breaks out of its parent MC, its shock wave would sweep up different amounts of molecular gas in different directions with different decelerations, resulting in an offset bubble structure.
Such a structure will be further enhanced by the radial density gradient of the MC.
According to the distribution of the velocity component in the line of sight of different parts of the shell, a similar structure will be present in the position-velocity map as shown in Figure~\ref{f:f1}.
The blowout morphology has been observed in many SNRs, e.g., 3C 400.2, RCW 103, and G352.7$-$0.1, which can be explained by the model in which the supernova explosion producing the remnant occurs in the medium with a density gradient or inside and near the border of a dense cloud \citep{Ambrocio-Cruz+2006, Toledo-Roy+2014, Lu+2021}.
Figure~\ref{f:12stamp} shows the distribution of \twCO\ (J=1--0) emission in the velocity range of $+$34 to $+$82 \km\ps\ with an interval of 3 \km\ps.

 

\end{appendix}
\end{document}